\documentstyle[twocolumn,aps,prd]{revtex}

\begin{document}
\draft
\preprint{}
\title{Multiple diffraction model for proton-proton elastic 
scattering\\
and total cross section extrapolations to cosmic-ray energies
}
\author{A. F.  Martini and M. J.  Menon}
\address{
Instituto de Fisica `Gleb Wataghin'\\
Universidade Estadual de Campinas, Unicamp\\
13083-970 Campinas, SP, Brasil
}
\date{\today}
\maketitle
\begin{abstract}
We analyse pp elastic scattering data at the highest accelerator 
energy region ($10 < \sqrt{s} \leq 62.5$ GeV) through a multiple 
diffraction approach. The use of Martin's formula in a model 
developed earlier is substituted by the introduction of a complex 
elementary (parton-parton) amplitude. With this the total cross 
section and the $\rho$-parameter may be simultaneously 
investigated and, with the exception of the diffraction minimum 
at some ISR energies, a satisfactory description of all 
experimental data is obtained. Total cross sections extrapolations 
to cosmic-ray energies ($\sqrt{s} > 6$ TeV) show agreement with the 
reanalysis of the Akeno data performed by N. N. Nikolaev and also 
with Gaisser, Sukhatme and Yodh results, leading to the prediction 
$\sigma_{tot}^{pp}(\sqrt{s}=16\;TeV)=147\;mb$. Physical
interpretations  and critical remarks concerning our 
parametrizations and results are also presented and discussed.
\end{abstract}

%\tableofcontents

%\widetext

\section{Introduction}
\label{sec:intro}

Elastic proton-proton scattering is the most simple process in 
high-energy hadronic interactions. Despite the amount of experimental 
data available and model descriptions of these data, a treatment 
based exclusively in the field theory of strong interactions, the 
quantum-chromodynamics (QCD), is still missing. Beyond the intrinsic 
interest coming from this fact, new expectations are associated 
with the next proton collider, the CERN Large Hadron Collider (LHC), 
optimistically planned to begin experiments at 14 TeV in 2005. The 
situation suggests that a fundamental and difficult task for present 
and future developments, is to find connections between model 
descriptions and reliable calculational schemes in QCD. Since 
elastic scattering incorporates soft processes one expects that 
nonperturbative formalisms will play a fundamental role.

Based on the above considerations we have investigated elastic 
proton-proton scattering through phenomenological approaches and, 
simultaneously, looking for connections with nonperturbative QCD 
treatments. With this strategy, we improved some aspects of a 
multiple diffraction model developed earlier. The main points concern 
the use of a complex elementary (parton-parton) amplitude instead of 
Martin's real-part formula and the selection of more suitable 
parametrizations for the free parameters. With this, we achieved a 
good (but limited) description of the main physical observables in 
the accelerator energy domain. Based on these satisfactory results we 
extrapolate our parametrizations to cosmic-ray energies in order to 
investigate total cross sections.

In this report we present in some detail all the underlying aspects 
of the work, related to both the technical matters and to the general 
ideas and physical interpretations.

The material is organized as follows. In Sec.\ \ref{sec:expe} we 
briefly recall the experimental status (accelerator and cosmic-ray 
data) on pp scattering and outline a survey of some theoretical 
results that show the present relevance of the multiple 
diffraction formalism. In Sec.\ \ref{sec:multi} we describe the 
model, the improvements introduced, the predictions for the physical 
observables in the accelerator energy region and the extrapolations 
to cosmic-ray energies. Discussions, physical interpretations and 
critical remarks concerning the main results are the content of 
Sec.\ \ref{sec:discu}.

\section{Experimental and theoretical contexts}
\label{sec:expe}

Elastic pp scattering is essentially characterized by three  
physical observables: differential cross section, total cross section 
and the $\rho$-parameter (ratio of the forward real to imaginary part 
of the scattering amplitude). From these, others quantities may be 
derived such as the slope parameter and integrated elastic and 
inelastic cross sections \cite{block85}.

Experimental data on these observables obtained and compiled nearly 
fifteen years ago, still remain the only source of information at 
high-energies extending up to $\sqrt{s}=62.5$ GeV, the CERN 
Intersecting Storage Ring (ISR) energy region 
\cite{amaldi80,schubert79}. Experimental information on pp total 
cross sections in the range of energy $\sqrt{s}: 6 - 40$ TeV 
exists from cosmic-ray data on Extensive Air Showers 
\cite{honda93,gaisser87,nikolaev93}. However, since the proton 
cross section is extracted from the proton-air cross section through phenomenological models \cite{gaisser87} the results are model 
dependent. Also, the analysis performed by the Akeno Collaboration \cite{honda93} was recently criticized by N.N. Nikolaev who claimed 
that the Akeno results have been underestimated by about 30 mb 
\cite{nikolaev93} and this is in agreement with the results early 
obtained by Gaisser, Sukhatme and Yodh \cite{gaisser87}. This is a 
central point in our work and we will discuss this discrepancy in 
Secs.\ \ref{sec:multi} and \ref{sec:discu}. Anyway, before the 
CERN's LHC proton-proton collider, cosmic-ray results remain the 
only source of information on pp interactions at the highest 
energies.

From the theoretical point of view it is well known that a pure QCD 
description of the elastic hadron scattering has not yet been 
obtained. Perturbative QCD cannot be extended to the soft region and 
pure nonperturbative QCD is not able to predict scattering states. 
Although the bulk of experimental data has been successfully 
described by different models in different contexts 
\cite{matthiae94}, a widely established and accepted approach is 
still missing. The same can be said of a pure QCD understanding of 
the subject.

As stated before, we are interested in the interface phenomenology/
nonperturbative QCD. For this we will limit this introductory 
discussion to some ideas directly related to this subject.

Despite the above difficulties, there has been in recent years 
important progresses in the framework of nonperturbative QCD, 
concerning soft processes. Looking for a microscopic approach able 
to explain phenomenological results, Landshoff and Nachtmann 
associated the Pomeron to the exchange of two abelian gluons with 
modified propagators \cite{landshoff87}. Following this analysis, 
Nacthmann extended the approach to the case of non-abelian gluons in 
a functional representation of scattering matrix elements and eikonal 
approximation \cite{nacthmann91}. With this, the quark-quark 
amplitudes are associated with a gluonic correlation function. 
However, since physical observables are connected with hadron-hadron 
amplitudes, a central problem concerns the construction of these 
amplitudes from the elementary ones.

One way to treat this problem is based on the stochastic vacuum model 
\cite{dosch87}. The hadronic amplitude is constructed through 
scattering amplitudes for Wilson loops in Minkowski space-time 
(loop-loop scattering), leading to gauge invariant amplitudes 
\cite{dosch92}. With this nonperturbative framework general 
characteristics concerning total cross sections and slope parameters 
in high-energy hadron-hadron scattering may be described 
\cite{dosch92}.

Another way to construct the hadronic amplitude from the elementary 
one is by means of Glauber's multiple diffraction theory 
\cite{glauber59} and this is the point we are interested in. The 
approach is based on the impact parameter and eikonal formalisms as 
follows: Assuming azimuthal symmetry in the collision of two hadrons 
A and B, the impact parameter formalism connects the eikonal $\chi$ 
and the elastic hadronic scattering amplitude by \cite{glauber59}

\begin{equation}
F(q,s)=i\int_0^{\infty} bdb[1 - e^{i\chi (b,s)}] J_{0}(qb) \equiv 
i<1 - e^{i\chi (b,s)}>,
\label{ampltrans}
\end{equation}
where $q^2=-t$ is the four-momentum transfer squared, $b$ the impact 
parameter, $J_{0}$ the zero-order Bessel function and the angular 
brackets denotes the symmetrical two-dimensional Fourier transform. 
In the first order Glauber multiple diffraction theory the eikonal is 
expressed as the Fourier transform of the products of the hadronic 
form factors, $G_A$ and $G_B$, by the averaged elementary 
(parton-parton) amplitude $f$, \cite{czyz69}, namely:

\begin{equation}
\chi (b,s)=<G_A G_B f>.
\label{eicotrans}
\end{equation}
This approximation means that at any time one constituent of a 
hadron interacts with only one constituent of the other hadron. 
This corresponds to the generalized form of the Chou-Yang model 
\cite{chou68}, once a well-defined Fourier transform for the 
elementary amplitude is assumed \cite{menon93}.

The importance of this phenomenological framework in the search for 
connections between experimental data and calculational schemes in 
nonperturbative QCD has been recently expressed in the work by 
Grandel and Weise \cite{grandel95}. Making use of the Dosch and 
Kr\"amer ansatz for the gluonic correlation function \cite{dosch92} 
the authors calculated the elementary (parton-parton) amplitude 
$f(q)$ in the high-energy approximation. Then, introducing a 
monopole parametrization for the unknown hadronic form factors in 
(\ref{eicotrans}) the hadronic amplitude (\ref{ampltrans}) could be 
calculated, leading to the differential cross section. With this, 
the authors obtained a satisfactory description of the differential 
cross section data for pp and \=pp elastic scattering at ISR and 
CERN Super Proton Synchroton (Sp\=pS) energies, in the region of 
small transfer momentum ($q^2\alt 2.0\;GeV^2$) \cite{grandel95}.

Despite all these developments we see that the connections between 
nonperturbative QCD approaches and the bulk of physical observables 
still depends strongly on phenomenological models and ad hoc 
parametrizations. In this sense we understand that the role of a 
constant feedback on phenomenological information is crucial for 
present and future developments.

In particular we see that the multiple diffraction formalism may 
be a powerful tool in the test of suitable parametrizations. 
Simultaneously it may contribute with the search of reliable 
calculational schemes at differents levels of the theory (e.g. the 
stochastic vacuum model). We stress also the importance of the energy 
dependences which may be extracted from parametrizations for the form 
factors and elementary amplitudes. Since the theoretical approaches 
(nonperturbative) mentioned above make use of asymptotic energy 
limits, the full increase of the total cross section, the shrinkage 
of the diffraction peak, the energy dependence of the 
$\rho$-parameter and even possible residual differences between pp 
and \=pp scattering at the highest energies have not yet been 
explained.

Based on all these considerations, in the next sections we will 
investigate elastic pp scattering in the context of a multiple 
diffraction model.

\section{Multiple diffraction model}
\label{sec:multi}
\subsection{Previous approach and Martin's formula}
\label{sec:multiprev}

From a phenomenological point of view, multiple diffraction models 
are currently identified by each particular choice of 
parametrizations for the form factors, $G_A, G_B$ and elementary 
amplitude, $f$ in Eq.\ (\ref{eicotrans}) \cite{menon93}. With 
this, the hadronic amplitude (\ref{ampltrans}) may be investigated 
and then, in principle, the physical observables referred to in 
the beginning of Sec.\ \ref{sec:expe}

\begin{equation}
{d\sigma\over dq^2}=\pi |F(q,s)|^2 ,
\label{scdiff}
\end{equation}

\begin{equation}
\sigma_{tot}(s)=4\pi Im\{F(q=0,s)\} ,
\label{sctot}
\end{equation}

\begin{equation}
\rho(s)=\frac{Re\{F(q=0,s)\}}{Im\{F(q=0,s)\}}.
\label{rhoparam}
\end{equation}

In a previous approach, elastic pp and \=pp scattering were 
investigated in a multiple diffraction model through the following 
parametrizations for the form factors and elementary amplitude 
\cite{menonpim93,menon92}

\begin{equation}
G=(1+\frac{q^2}{\alpha^2})^{-1}(1+\frac{q^2}{\beta^2})^{-1}
\label{fatorform}
\end{equation}

\begin{equation}
f=iC\frac{1-\frac{q^2}{a^2}}{1+\frac{q^4}{a^4}}
\label{amplelem}
\end{equation}
where $\alpha^2, \beta^2, a^2$ and $C$ are free parameters. We 
introduced here a small change in the previous notation of the 
elementary amplitude which will be suitable in what follows.

The justifications for the above parametrizations were 
extensivelly discussed and explained in \cite{menon92} and 
\cite{menon96} and will not be reproduced here. We only recall 
that Eq.\ (\ref{amplelem}) is in agreement with the results from 
model-independent analysis for the 
dynamical part of the eikonal as obtained by Buenerd, Furget, and 
Valin \cite{furget90} and Carvalho and Menon \cite{carvalho97}. 
Discussion on this subject may be found in \cite{menon93} and 
\cite{martini96}.

With the parametrizations (\ref{fatorform}) and (\ref{amplelem}) the 
eikonal in (\ref{eicotrans}) is purelly imaginary and so the hadronic 
amplitude (\ref{ampltrans}). In the previous approach the real part 
of this amplitude was estimated through Martin's formula 
\cite{martin73}

\begin{equation}
Re F(q,s)=\frac{d}{dq^2}[\rho q^2 Im F(q,s)],
\label{martinform}
\end{equation}
using the experimental $\rho$ value at each energy. With this 
approach a satisfactory description of experimental data on 
differential and integrated cross sections was obtained 
\cite{menonpim93,menon96,menonhad93}.

However, a crucial point concerns the use of the above formula and 
this deserves some discussion. First, as derived by Martin the 
formula holds only for values of the momentum transfer 
infinitesimally small and at the asymptotically high-energy regime 
\cite{martin73}. The formula may also be derived through the 
geometrical scaling hypothesis \cite{diasdedeus73} and in this case 
its applicability should be limited to the ISR energy region. 
Corrections to Martin's formula were introduced by Henzi and Valin 
\cite{henzi84} and numerical analysis from fits of experimental data 
by Kundr\'{a}t and Lokajic\u{e}k puts serious limits in its 
applicability concerning momentum transfer \cite{kundrat85}. This 
result however has been recently criticized by Kawasaki, Maehara and 
Yonezawa who present results favouring the applicability of the 
formula in the whole region of the momentum transfer with the data 
available \cite{kawasaki96}.

Despite this controversial aspect, a serious problem with the use 
of Martin's prescription is that $\rho(s)$ is an input parameter at 
each energy and so cannot be investigated. This led us to try a 
different procedure in the determination of the real part of the 
hadronic amplitude and in the next section we introduce a possible 
solution for the problem.

\subsection{Complex elementary amplitude}
\label{sec:multicomp}

In the context of the multiple diffraction formalism, Eqs.\ 
(\ref{ampltrans}) and (\ref{eicotrans}), associated with a complex 
hadronic amplitude, $F_{AB}$, one should expect a complex elementary 
(parton-parton) amplitude

\[ f(q,s)=Re f(q,s) + i Im f(q,s) . \]
In this sense, the approach discussed in the last section corresponds 
to the assumption of $Re f(q,s)=0$ and

\begin{equation}
Im f(q,s)= C\frac{1-\frac{q^2}{a^2}}{1+\frac{q^4}{a^4}} .
\label{amplelemim}
\end{equation}
Lacking both theoretical and experimental information about the 
elementary phase we will assume, as a first approximation, a 
proportionality relation between real and imaginary parts at each 
energy \cite{martini95}

\begin{equation}
Re f(q,s)=\lambda(s) Im f(q,s) ,
\label{amplelemre}
\end{equation}
where $\lambda(s)$ is a free parameter. With this assumption the 
eikonal in (\ref{eicotrans}) may be expressed by

\begin{equation}
\chi(b,s)= (\lambda+i)\Omega(b,s)
\label{eicoopac}
\end{equation}
where, for the proton-proton case,

\begin{equation}
\Omega(b,s)=<G^2 Im f(q,s)> ,
\label{omegapp}
\end{equation}
with $G$ given by (\ref{fatorform}) and $Im f(q,s)$ by 
(\ref{amplelemim}). With this, the real and imaginary parts of the 
hadronic amplitude read

\begin{equation}
Re\{F(q,s)\}=<e^{-\Omega(b,s)}\sin(\lambda\Omega(b,s))> ,
\label{amplre}
\end{equation}

\begin{equation}
Im\{F(q,s)\}=<1-e^{-\Omega(b,s)}\cos(\lambda\Omega(b,s))>.
\label{amplim}
\end{equation}

Substituting parametrizations (\ref{fatorform}) and 
(\ref{amplelemim}) for $G$ and $Im f$ in (\ref{omegapp}), the 
``opacity'' $\Omega(b,s)$ is analytically evaluated. Then, 
Eqs.\ (\ref{amplre}) and (\ref{amplim}) lead to the differential 
cross section (\ref{scdiff}).

As a first test of the ansatz (\ref{amplelemre}) we calculate the 
imaginary part of the hadronic amplitude, using yet the 
parametrizations from the previous approach \cite{menonpim93,menon92}
; then the real part was evaluated through both, Martin's 
prescription (\ref{martinform}) and the ansatz (\ref{amplelemre}). 
For pp elastic scattering at $\sqrt{s}=52.8$ GeV (largest interval in 
momentum transfer with data available, \cite{schubert79}) the 
differential cross section is well described for $\lambda=0.055$ and 
we display both results in Fig.~\ref{fig1}.

We see that the predictions for the real part are similar in both 
cases: they present two zeros (change of sign) and its contributions 
to the differential cross sections are important only in the dip 
region.

Since the imaginary part of the amplitude presents a zero and the 
real part of the Martin formula is obtained by the derivative 
(\ref{martinform}), the contribution of this part is out of phase and 
then the differential cross sections does not vanish in the dip 
region \cite{martin73}. With the assumption (\ref{amplelemre}) for 
the elementary amplitude the same effect is obtained in the hadronic 
amplitude due to $\sin$ and $\cos$ terms in (\ref{amplre}), 
(\ref{amplim}), and the result is in qualitative agreement with the 
predictions of zeros (change of sign) from dispersion relation 
\cite{bronzan74}.

Based on this satisfactory result with the complex elementary 
amplitude, we investigated the simultaneous descriptions of all 
physical observables referred before, Eqs.\ (\ref{scdiff}), 
(\ref{sctot}) and (\ref{rhoparam}). To our knowledge this has never 
been achieved through geometrical or multiple diffraction models 
and we will return to this point later.

Trying global descriptions for pp elastic scattering, besides the 
use of the complex elementary amplitude, we reanalysed the fits and 
parametrizations, improving some aspects of the previous approach. We 
discuss in detail the central points in what follows.

\subsection{Fits of experimental data}
\label{sec:multifits}

As explained in the last two sections, our approach has only five 
free parameters: two associated with the form factor, $\alpha^2$ and 
$\beta^2$, and three with the elementary amplitude, $C, a^2$ and 
$\lambda$.

We analysed 7 sets of pp experimental data above 10 GeV (Table 
\ref{tabparam}) and the fits were performed only of the differential 
cross section data \cite{schubert79,ayres77} and $\rho$ parameter 
data \cite{amaldi80,fajardo81} at each energy. The fit procedure 
consists of two steps:

\begin{enumerate}
  \item Taking $\lambda=0$ in Eq.\ (\protect\ref{eicoopac}) the 
hadronic amplitude, Eqs.\ (\protect\ref{amplre}) and 
(\protect\ref{amplim}), is purely imaginary. For this case we 
determined the values of the parameters $C, \alpha^2, \beta^2$ and 
$a^2$ that reproduce the differential cross section data at each 
energy so as to present the zero at the dip position.
  \item With the values of the above four parameters as input we 
then calculated the value of $\lambda$ that reproduces the 
experimental $\rho$ value at each energy.
\end{enumerate}
Step 1 comes essentially from the previous approach and has been 
discussed and explained in detail in \cite{menon96}.

With the above procedure a satisfactory description of $\rho$ and 
$d\sigma/dq^2$ experimental data was obtained with two constant free 
parameters

\begin{equation}
a^2=8.20\;GeV^2,\quad\quad \beta^2=1.80\;GeV^2, 
\label{abeta}
\end{equation}
and only three parameters depending on the energy: $C(s),\alpha^2(s)$ 
and $\lambda(s)$. The values are shown in Table \ref{tabparam} for 
each set analysed.

In order to obtain a formalism able to make predictions to other 
energies we then proceed to investigate parametrizations for the data 
displayed in Table \ref{tabparam}.

\subsection{Parametrizations as a function of the energy}
\label{sec:multipara}

The choice of suitable and consistent parametrizations is a crucial 
point, mainly concerning extrapolations. We will discuss this 
aspect in detail in what follows and also in Sec.\ \ref{sec:discu}. 
We first analyze the dependences of $C$ and $\alpha^{-2}$, which 
determines the imaginary part of the hadronic amplitude (and so the 
total cross section) and after the dependence of $\lambda$ 
(associated only with the real part of the amplitude).

The values of $C$ and $\alpha^{-2}$ from Table \ref{tabparam} are 
displayed in Fig.~\ref{fig2} and we see that both increase with the 
energy, presenting positive curvatures.

Experimentally, total cross sections grow like $[\ln s]^n$, $n=1$ or 
2, at and above ISR and there is indication of ``qualitative'' 
saturation of the Martin-Froissart bound, $n=2.2\pm 0.3$ 
\cite{augier93,matthiae94}. Also, from gauge field theory, lowest 
order cross sections for particle production (unitarity) present 
$\ln s$ terms \cite{gotsman93,cheng87}. Based on these facts and from 
the behaviour shown in Fig.~\ref{fig2}, we introduced fits through 
polynomials

\[\sum_{n=0}^N a_n [\ln \frac{s}{s_0}]^n , \]
which is different from our early parametrizations 
\cite{menonpim93,menon92,menon96}. Both sets of points are 
statistically consistent with polynomials of second degree 
\cite{bevington92} and through linear regression we obtained 
\cite{martinime96}

\begin{equation}
C(s)=14.3-1.65[\ln (s)]+0.159[\ln (s)]^2 , \quad (GeV^{-2})
\label{clns}
\end{equation}

\begin{equation}
{1\over \alpha^2}=2.57-0.217[\ln (s)]+0.0243[\ln (s)]^2 , 
\quad (GeV^{-2})
\label{alfalns}
\end{equation}
and we took $s_0=1\;GeV^2$. In Fig.~\ref{fig2} we show the above 
parametrizations together with the fit values from Table 
\ref{tabparam}. Physical interpretations and critical remarks 
concerning parametrizations (\ref{clns}) and (\ref{alfalns}) will be 
presented in Sec.\ \ref{sec:discublac}.

In the case of parameter $\lambda$ the choice of a suitable 
parametrization demands further discussions. Empirical analysis of 
the influence of $\lambda$ in the hadronic amplitude showed that its 
behaviour is similar to that of the $\rho$ parameter. That is, if 
$\lambda$ increases (decreases) also $\rho$ increases (decreases) and 
$\lambda=0$ at the same energy value where $\rho=0$. We will return 
to this point in Sec.\ \ref{sec:discureal}. Since there is no 
experimental information on $\rho(s)$ above 62.5 GeV we have, in 
principle, serious limitations in the choice of parametrizations. 
That is, in order to make extrapolations to high energies in our 
strictly phenomenological approach, we should infer the $\rho$ 
behaviour or investigate limiting cases. With this last possibility 
in mind, we recall the general belief that, above $\sqrt{s}\sim 100$ 
GeV, $\rho(s)$ has a maximum and then goes asymptotically to zero 
through positive values \cite{matthiae94}. However, how fast this 
happens depends on model assumptions. In order to test these 
possibilities we considered two different approaches to zero, which 
could be understood as some kind of limiting cases, that is, slow and 
fast convergences. The point as we shall show is to investigate the 
influence of these assumptions in the description of the 
experimental data.

In Fig.~\ref{fig3} we display the values of $\lambda$ from Table 
\ref{tabparam}. Based on this behaviour and on the above 
considerations we introduce the following general parametrization:

\begin{equation}
\lambda(s)={A_1ln(s/s_0)\over 1+A_2[ln(s/s_0)]+A_3[ln(s/s_0)]^2} .
\label{lambdalns}
\end{equation}
In this formula $s_0$ controls the point where $\lambda(s)$ (and 
$\rho(s)$) reaches zero and $A_i,\; i=1,2,3$ the maximum and 
asymptotic behaviour. As the two limiting cases we took:

\begin{eqnarray}
Case\;1: & A_1=6.95\times 10^{-2},\>A_2=0.118,\nonumber \\
& A_3 = 1.50\times 10^{-2}
\label{lambdacaso1}
\end{eqnarray}

\begin{eqnarray}
Case\;2: & A_1=9.08\times10^{-2},\>A_2=0.318,\nonumber \\ 
& A_3=1.70\times 10^{-10}
\label{lambdacaso2}
\end{eqnarray}
and $s_0=400\;GeV^2$ in both cases. Figure \ref{fig3} shows these 
parametrizations up to $\sqrt{s}=10^5\;GeV$.

With the parametrizations for $C(s), \alpha^{-2}(s)$ and $\lambda(s)$
, Eqs.\ (\ref{clns}), (\ref{alfalns}) and (\ref{lambdalns}), 
respectively (cases 1 and 2), and result (\ref{abeta}) for the 
remaining parameters $\beta^2$ and $a^2$, all free parameters are 
completely determined. Through the formalism described in Secs.\ 
\ref{sec:multiprev} and \ref{sec:multicomp}, Eqs. (\ref{fatorform}), 
(\ref{amplelemim}) and (\ref{omegapp}) to (\ref{amplim}), the three 
physical observables (\ref{scdiff}), (\ref{sctot}) and 
(\ref{rhoparam}) may be predicted. 

\subsection{Model predictions and experimental data}
\label{sec:multimode}

As explained in Sec.\ \ref{sec:multifits} our fits were performed 
only on differential cross section and $\rho$ parameter data in the 
interval $\sqrt{s}=13.8 - 62.5$ GeV. In this session we first check 
the predictions in the interval $\sqrt{s}=10 - 100$ GeV (accelerator 
energy region) and then display the extrapolations up to $10^5$ GeV 
(cosmic-ray energies and future accelerators).
\begin{itemize}
  \item Accelerator energy region
\end{itemize}

Figures \ref{fig4} and \ref{fig5} show the model 
predictions (cases 1 and 2) for the pp differential cross section, 
$\rho$ parameter and total cross sections, together with the 
experimental data available. We observe that in this interval, cases 
1 and 2 are distinguishable only for $\rho(s)$ and 
$\sigma_{tot}^{pp}(s)$ and below $\sqrt{s}\sim 15$ GeV. With the 
exception of the diffraction minimum (dip) at the highest ISR 
energies, the agreement with the experimental data is quite good. 
In Sec.\ \ref{sec:discureal} we will return to this overestimation 
of the differential cross section in the dip region.
\begin{itemize}
  \item Extrapolations to higher energies
\end{itemize}

Assuming that our parametrizations hold at higher energies we 
calculate the predictions for $\rho(s)$, $\sigma_{tot}^{pp}(s)$ up to 
$\sqrt{s}=10^2$ TeV (cosmic-ray region) and for the differential 
cross sections at 10, 15 and 20 TeV (future LHC).

We show in Fig.~\ref{fig6} the predictions for $\rho(s)$ in 
cases 1 and 2 (Fig.~\ref{fig5}) and the experimental data 
available. We understand that case 1 should be the most reliable from 
a conservative point of view. The similarities between $\rho(s)$ and 
$\lambda(s)$ (Fig.~\ref{fig3}) will be discussed in Sec.\ 
\ref{sec:discureal}.

The predictions for the total cross section are shown in 
Fig.~\ref{fig7}, together with accelerator data 
(Fig.~\ref{fig5}) and results from cosmic-ray experiments 
which, due to the existing discrepancies, we briefly review in what 
follows.

The information available on proton-proton total cross section, 
$\sigma_{tot}^{pp}$, from cosmic-ray air showers are obtained through 
the p-air inelastic cross section, $\sigma_{p-air}^{inel}$. However, 
either the determination of the $\sigma_{p-air}^{inel}$ or the 
relation between $\sigma_{p-air}^{inel}$ and $\sigma_{tot}^{pp}$ are 
model dependent \cite{gaisser87}. In the detailed analysis of data 
from Fly's eyes experiment, Gaisser, Sukhatme and Yodh (GSY 
hereafter), estimated the limit

\[ \sigma_{tot}^{pp} \geq 130\;mb \; at \; \sqrt{s}\sim 30\;TeV. \]
Making use of the Chou-Yang relation between $\sigma_{tot}^{pp}$ 
and the slope parameter, they calculated \cite{gaisser87}

\[ \sigma_{tot}^{pp}=175_{-27}^{+40}\;mb \; at \; \sqrt{s}=40\;TeV.\]
More recently, based on analysis of the extensive air shower, the 
Akeno Collaboration presented results in the interval 
$\sqrt{s}: 6 - 25$ TeV. In particular they found \cite{honda93} 

\[ \sigma_{tot}^{pp}=133\pm 10\;mb \; at \; \sqrt{s}=40\;TeV, \]
which is incompatible with the GSY result. However, Nikolaev showed 
that $\sigma_{p-air}^{inel}$ inferred by the Akeno group  should be 
identified with an absorption cross section and that this originates 
an increase of $\sim 30\;mb$ in the Akeno results for 
$\sigma_{tot}^{pp}$. From the Nikolaev analysis \cite{nikolaev93},

\[ \sigma_{tot}^{pp}=160 - 170\;mb \; at \; \sqrt{s}=40\;TeV, \]
which is in agreement with the GSY calculation.

Although the Akeno results are usually referred to in the literature, 
the analysis either by Nikolaev or by GSY seems correct and both are 
incompatible with Akeno.

All these informations are displayed in Fig.~\ref{fig7}, 
together with accelerator data and our predictions in cases 1 and 2. 
We see that the model predictions are in complete agreement with 
Nikolaev and GSY results and this implies in a faster increase of 
the total cross section than usually believed. In particular we 
predict

\[ \sigma_{tot}^{pp}=147\;mb \; at \; \sqrt{s}=16\;TeV,  \]
and

\[ \sigma_{tot}^{pp}=176\;mb \; at \; \sqrt{s}=40\;TeV.  \]
Further discussion on the subject will be presented in Sec.\
 \ref{sec:discuhigh}.

Concerning the differential cross section, we calculate the 
predictions in the region to be reached by the CERN's LHC, which are 
displayed in Fig.~\ref{fig8}. The results present no dip but a 
shoulder at $|t|\sim 0.5\;GeV^2$ and at $|t|>4\;GeV^2$ the curves 
decrease smoothly with no other structures.

\section{Discussion}
\label{sec:discu}
The model described presents three parameters that depend on  
energy: $\alpha^2, C$ and $\lambda$. In this last section we discuss 
some critical points and physical interpretation concerning the 
parametrizations and predictions.

\subsection{High-energy extrapolations and total cross sections}
\label{sec:discuhigh}

Our parametrizations (\ref{clns}) and (\ref{alfalns}) for $C(s)$ and 
$\alpha^{-2}(s)$, respectively, were based on experimental 
information only in the interval $13.8 \leq \sqrt{s} \leq 62.5$ GeV 
and in both cases the points indicate an increase with the energy 
(Table \ref{tabparam}, Fig.~\ref{fig2}). A crucial question concerns 
this limited region since different explicit parametrizations, 
statistically consistent with the set of points, may differ 
arbitrarily when extrapolated to higher energies.

However, constraints on the choice of parametrizations may be found 
through physical information available and this played an important 
role in our approach. As mentioned in Sec.\ \ref{sec:multipara}, our 
strategy was based on the general assumption of the expected $\ln s$ 
-- behaviour of soft processes and on the reasonable hypothesis of 
polynomial functions on $\ln s$. With these constraints, we have a 
linear system in the free parameters of the polynomials (with the 
exception of the assumed value $s_0=1$ GeV) and the statistical 
solution is unique \cite{bevington92}, leading to the forms 
(\ref{clns}) and (\ref{alfalns}) which describe the points quite 
well.

On extrapolating the predictions for the total cross section to 
cosmic-ray energies, our results agree very well with both  
analyses by Nikolaev \cite{nikolaev93} and by Gaisser, Sukhatme 
and Yodh \cite{gaisser87} (Fig.~\ref{fig7}). Since the 
approaches by these authors are totally independent of the 
considerations and assumptions we have made, and, as far as we 
know, there is no criticism concerning their results, the agreement 
shown in Fig.~\ref{fig7} suggests a real increase of the pp 
total cross section faster than generally expected. In particular 
we predict 

\[ \sigma_{tot}^{pp}=91.6\;mb \; at \; \sqrt{s}=1.8\;TeV, \]
which is higher than even the CDF result for \=pp, 
$\sigma_{tot}^{\bar{p}p}=80.03\pm 2.24\;mb$ \cite{giromini94}. This 
seems to favour the ``Odderon hypothesis'' \cite{lukaszuk73}, a 
problem which still ``remains entirely open both from the theoretical 
as well as from the experimental point of view'' \cite{giffon96}.

\subsection{Blackening and expansion}
\label{sec:discublac}

The parameter $\alpha^2$ coming from the hadronic form factor 
(\ref{fatorform}) is associated with the radius defined by 

\[ R^2(s)=-6 \frac{dG(q,s)}{dq^2}|_{q^2=0}  \]
and from (\ref{fatorform})

\begin{equation}
R(s)=(0.483) [{1\over \alpha^2(s)}+{1\over \beta^2}]^{1\over 2}\; 
(fm).
\label{raioalfa}
\end{equation}

Through parametrization (\ref{alfalns}) for $\alpha^{-2}$ and 
$\beta^2=1.80\;GeV^2$, Eq. (\ref{abeta}), the radius increases with 
the energy as shown in Fig.~\ref{fig9}. We can then interpret the 
parameter $\alpha^2$ as associated with the well known ``expansion 
effect'' \cite{cheng87}.

The parameter $C$ corresponds to the ``absorption constant'' in the 
Chou-Yang picture \cite{chou68,chou67} and is associated with the 
number of constituent partons in the context of the Glauber 
approach \cite{glauber59,czyz69}. It then controls the ``blackening 
effect'' coming from the absorption. Our results, as in the previous 
approach \cite{menonpim93,menon96}, mean that hadrons become blacker 
and simultaneously larger as the energy increases, in agreement with 
the ``BEL behaviour'' (Black, Edgie and Large) found by Henzi and 
Valin \cite{henzi83}.

These effects may be investigated through the behaviour of the 
Inelastic Overlap Function, $G_{in}(b,s)$, which is calculated from 
the unitarity condition in the impact parameter space \cite{amaldi76}

\begin{equation}
2Re \Gamma (b,s) = |\Gamma (b,s)|^2 + G_{in} (b,s) ,
\label{ginel}
\end{equation}
where $\Gamma (b,s)$ is the Profile Function, the Fourier transform 
of the hadronic amplitude (\ref{ampltrans}):

\begin{equation}
\Gamma (b,s) = 1 - e^{i\chi (b,s)} .
\label{profile}
\end{equation}

In Fig.~\ref{fig10} we show $G_{in} (b,s)$ as function of the energy 
and for some fixed values of the impact parameter $b$. In the region 
10 to 100 GeV (part (a)) we observe the simultaneous increase of 
$G_{in}$ at all values of the impact parameter. In part (b) we show 
the central ($b=0$) and peripheral ($b=1 fm$) regions extrapolated to 
$\sqrt{s}=10^5$ GeV. We observe that in the central region the 
curvature becomes negative above $\sqrt{s}=100\sim 200$ GeV and in 
the peripheral region, above $\sqrt{s}\sim 10^3$ GeV. The black disc 
limit ($G_{in}=1$) however seems very far to be reached, i.e., 
much higher than $10^5$ GeV.

Since $C(s)$ and $\alpha^2(s)$ control the blackening and expansion 
effects, respectively, the dimensionless quantity $C\alpha^2$ gives 
information on the influence of each effect as function of the 
energy. Figure~\ref{fig11} shows the predictions obtained by 
means of parametrizations (\ref{clns}) and (\ref{alfalns}) up to 
$10^5$ GeV, together with the values from fit (Table \ref{tabparam}). 
We observe a minimum at $\sqrt{s}\sim 30$ GeV, a change of sign in 
the curvature (becomes negative) above $\sqrt{s}\simeq 10^3\;GeV$ 
and the asymptotic limit value $\sim 6.5$ (from the 
parametrizations).

\subsection{Forward real-to-imaginary ratios of the partonic and 
hadronic amplitudes}
\label{sec:discureal}

An essential characteristic of the Glauber multiple diffraction 
formalism is to connect elastic scattering cross sections for 
composite particles (originally nuclei and after nucleons) with the 
scattering amplitudes of their individual components (originally 
nucleons and after partons, respectively) 
\cite{glauber59,czyz69,glauber84}. In this context, the assumption 
(\ref{amplelemre}) of proportionality between real and imaginary 
parts of the elementary (parton-parton) amplitude means that, in 
particular,

\[ \lambda(s)=\frac{Re f(q=0,s)}{Im f(q=0,s)} , \]
i.e., at the partonic level $\lambda(s)$ plays the same role as 
$\rho(s)$ at the hadronic level, Eq.\ (\ref{rhoparam}). The 
similarities referred to in Sec.\ \ref{sec:multipara} between 
$\lambda(s)$ and $\rho(s)$ may be seen by comparisons of 
Figs.~\ref{fig3} and \ref{fig6} (cases 1 and 2).

The hypothesis that $\lambda$ does not depend on the momentum 
transfer is a very simple one and has been used here only as an 
ansatz. Despite this, from Fig.~\ref{fig1} the resulting 
contribution of the real part of the hadronic amplitude to the 
differential cross section is in qualitative agreement with the 
results obtained through Martin's prescription.

On the other hand, in the context of our approach, the limitations of 
this simple ansatz appear when we try simultaneous descriptions of 
cross sections (differential and total) and the $\rho$ parameter. As 
we show in Fig.~\ref{fig4}, the predictions overestimate the 
differential cross section in the dip region at the highest ISR 
energies, leading to a limited description of the set of physical 
observables. With the exception of this point, all the predictions 
are in agreement with the experimental data (Figs.~\ref{fig4} 
and \ref{fig5}). Since the total cross section is 
calculated from the imaginary part of the hadronic amplitude our 
novel results at cosmic-ray energies are independent of the effect in 
the dip region.

We observe that the ansatz (\ref{amplelemre}) may be formally 
equivalent to some other geometrical and multiple diffraction models 
characterized by complex eikonals 
\cite{furget90,glauber84,kamran88,saleem88}. The novel point here was 
to treat this hypothesis explicitly and investigate its consequence 
in the context of the Glauber approach, as, for example, the strong 
correlation between $\rho(s)$ and $\lambda(s)$ seen in Figs.
~\ref{fig3} and \ref{fig6}.

To our knowledge, simultaneous and complete descriptions of cross 
sections (differential and total) and the $\rho$ parameter, still 
remain an open problem in geometrical and multiple diffraction 
models. For this reason, we hope, our results may bring insights for 
further and deeper developments.

\acknowledgments

We are thankful to Capes and CNPq for financial support.

\begin{figure}
\caption{Results for the elastic pp differential 
cross section at $\protect\sqrt{s}=52.8$ GeV. The imaginary part of 
the amplitude was calculated using previous parametrizations 
\protect\cite{menonpim93,menon92} and the real part through (a) 
Martin prescription (\protect\ref{martinform}), and (b) the ansatz 
(\protect\ref{amplelemre}), with $\lambda=0.055$. In both cases it is 
shown the contributions from the real part (dotted), imaginary part 
(dashed) and complex amplitude (solid). Experimental data on 
$d\protect\sigma/dt$ are from \protect\cite{schubert79} and 
the $\rho$ value in Martin's formula from 
\protect\cite{amaldi80}.}
\label{fig1}
\end{figure}

\begin{figure}
\caption{Values of the free parameters $C$ and $\alpha^{-2}$, from 
Table \protect\ref{tabparam}, that fit experimental data (circles) 
and parametrizations through (\protect\ref{clns}) and 
(\protect\ref{alfalns}) (solid line).}
\label{fig2}
\end{figure}

\begin{figure}
\caption{Values of the free parameter $\protect\lambda$, from 
Table \protect\ref{tabparam}, that fit experimental data (circles) 
and parametrizations through (\protect\ref{lambdalns}) in the cases 
(\protect\ref{lambdacaso1}) (dot) and (\protect\ref{lambdacaso2}) 
(dash).}
\label{fig3}
\end{figure}

\begin{figure}
\caption{Model predictions for the differential cross sections in 
cases 1, Eq.\ (\protect\ref{lambdacaso1}), and 2, Eq.
\ (\protect\ref{lambdacaso2}) (indistinguishable) and experimental 
data (\protect\cite{ayres77} for $\protect\sqrt{s}=13.8$ and 19.4 
GeV and \protect\cite{schubert79} for the other energies). Curves 
and data were multiplied by factors of $10^{\protect\pm 2}$.}
\label{fig4}
\end{figure}

\begin{figure}
\caption{Model predictions for the $\protect\rho$ parameter (left) 
and total cross section (right) in cases 1, Eq.\ 
(\protect\ref{lambdacaso1}) (solid), and 2, Eq.\ 
(\protect\ref{lambdacaso2}) (dash) and experimental data 
(\protect\cite{amaldi80} and \protect\cite{fajardo81} for 
$\protect\rho$ and \protect\cite{amaldi80} and 
\protect\cite{carrol76} for the total cross section).}
\label{fig5}
\end{figure}

\begin{figure}
\caption{Predictions for the $\rho$ parameter in cases 1 and 2 and 
experimental data (Fig.~\protect\ref{fig5}).}
\label{fig6}
\end{figure}

\begin{figure}
\caption{Predictions for the proton-proton total cross section and 
experimental informations: accelerator data 
\protect\cite{amaldi80,carrol76} (crosses), Akeno 
\protect\cite{honda93} (circles), Nikolaev \protect\cite{nikolaev93} 
(triangles), GSY limit at 30 TeV \protect\cite{gaisser87} ($\uparrow$) 
and GSY result at 40 TeV \protect\cite{gaisser87} (square).}
\label{fig7}
\end{figure}

\begin{figure}
\caption{Predictions for the pp differential cross section at 10 
(solid), 15 (dash) and 20 TeV (dot).}
\label{fig8}
\end{figure}

\begin{figure}
\caption{Radius calculated through Eq.\ (\protect\ref{raioalfa}) 
with $\protect\beta^2=1.80\;GeV^2$ and parametrization 
(\protect\ref{alfalns}) for $\protect\alpha^{-2}$.}
\label{fig9}
\end{figure}

\begin{figure}
\caption{Inelastic Overlap Function calculated from the eikonal 
through Eqs. (\protect\ref{ginel}) and (\protect\ref{profile}).}
\label{fig10}
\end{figure}

\begin{figure}
\caption{Predictions for the dimensionless quantity 
$C\protect\alpha^2$ (see text) and points from fit (Table 
\protect\ref{tabparam}).}
\label{fig11}
\end{figure}

\begin{table}
\caption{Values of the free parameters from fits of pp diferential 
cross section and $\rho$ data at each energy.}
\begin{tabular}{ccccc}
$\protect\sqrt{s}$& C(s)& $\alpha^{-2}$(s)& $\lambda$(s)& 
C$\alpha^2$\\
($GeV$)& ($GeV^{-2}$)& ($GeV^{-2}$)& & \\
\tableline
13.8& 9.970& 2.092& -0.094& 4.77 \\
19.4& 10.050& 2.128& 0.024& 4.72 \\
23.5& 10.250& 2.174& 0.025& 4.71 \\
30.7& 10.370& 2.222& 0.053& 4.67 \\
44.7& 10.890& 2.299& 0.079& 4.74 \\
52.8& 11.150& 2.370& 0.099& 4.70 \\
62.5& 11.500& 2.439& 0.121& 4.72 \\
\end{tabular}
\label{tabparam}
\end{table}


\begin{references}
\bibitem{block85}M.M. Block and R.N. Cahn, Rev. Mod. Phys. {\bf 57}, 
563 (1985).
\bibitem{amaldi80}Experimental data on $\rho$ and total cross 
sections between 23.5 and 62.5 GeV: U. Amaldi and K.R. Schubert, 
Nucl. Phys. {\bf B166}, 301 (1980).
\bibitem{schubert79}Differential cross section data between 23.5 and 
62.5 GeV: K.R. Schubert, ``Tables on nucleon-nucleon scattering'', 
{\it in} Landolt-B\"ornstein, Numerical Data and Functional 
Relationships in Science and Technology, New Series, Vol. I/9a 
(1979).
\bibitem{honda93}M. Honda {\it et al}., Phys. Rev. Lett. {\bf 70}, 
525 (1993).
\bibitem{gaisser87}T.K. Gaisser, U.P. Sukhatme, and G.B. Yodh, Phys. 
Rev. D{\bf 36}, 1350 (1987).
\bibitem{nikolaev93}N.N. Nikolaev, Phys. Rev. D{\bf 48}, R1904 
(1993).
\bibitem{matthiae94}G. Matthiae, Rep. Prog. Phys. {\bf 57}, 743 
(1994).
\bibitem{landshoff87}P. Landshoff and O. Nachtmann, Z. Phys. 
C{\bf 35}, 405 (1987).
\bibitem{nacthmann91}O. Nachtmann, Ann. Phys. (N.Y.) {\bf 209}, 436 
(1991).
\bibitem{dosch87}H.G. Dosch, Phys. Lett. B{\bf 190}, 177 (1987);
H.G. Dosch and Yu.A. Simonov, Phys. Lett. B{\bf 205}, 339 (1988).
\bibitem{dosch92}H.G. Dosch, E. Ferreira, and A. Kr\"amer, Phys. 
Lett. B{\bf 289}, 153 (1992); Phys. Lett. B{\bf 318}, 197 (1993); 
Phys. Rev. D{\bf 50}, 1992 (1994).
\bibitem{glauber59}R.J. Glauber, in {\it Lectures in Theoretical 
Physics}, edited by W.E. Britten {\it et al}. (Interscience, New 
York, 1959), Vol. I, p. 315; {\it High Energy Physics and Hadron 
Structure}, edited by S. Devons {\it et al}. (Plenum, New York, 
1970), p. 207.
\bibitem{czyz69}W. Czy\.z and L. C. Maximon, Ann. Phys. (N.Y.) 
{\bf 52}, 59 (1969); V. Franco and G. K. Varma, Phys. Rev. C
{\bf 18}, 349 (1978).
\bibitem{chou68}T.T. Chou and C.N. Yang, Phys. Rev. {\bf 175}, 1832 
(1968).
\bibitem{menon93}M.J. Menon, Phys. Rev. D{\bf 48}, 2007 (1993).
\bibitem{grandel95}U. Grandel and W. Weise, Phys. Lett. B{\bf 356}, 
567 (1995).
\bibitem{menonpim93}M.J. Menon and B.M. Pimentel, Hadronic J. 
{\bf 16}, 137 (1993).
\bibitem{menon92}M.J. Menon, Nucl. Phys. B (Proc. Suppl.) {\bf 25}, 
94 (1992).
\bibitem{menon96}M.J. Menon, Canadian J. Phys. {\bf 74}, 594 (1996).
\bibitem{furget90}C. Furget, M. Buenerd, and P. Valin, Z. Phys C
{\bf 47}, 377 (1990).
\bibitem{carvalho97}P.A.S. Carvalho and M.J. Menon, in {\it 1996 
XVII Brazilian National Meeting on Particles and Fields}, Serra 
Negra, SP, 1996 (to be published); Report No. IFGW ABSTRACTA C   
,1996.
\bibitem{martini96}A.F. Martini, M.J. Menon, and D.S. Thober, Phys. 
Rev. D{\bf 54}, 2385 (1996).
\bibitem{martin73}A. Martin, Lett. Nuovo Cimento {\bf 7}, 811 (1973).
\bibitem{menonhad93}M.J. Menon,  Hadronic J. {\bf 16}, 47 (1993).
\bibitem{diasdedeus73}J. Dias de Deus, Nucl. Phys. {\bf B59}, 231 
(1973); Il Nuovo Cimento {\bf 28}, 114 (1975); A.J. Buras and  
J. Dias de Deus, Nucl. Phys. {\bf B71}, 481 (1974).
\bibitem{henzi84}R. Henzi and P. Valin, Phys. Lett. B{\bf 149}, 239 
(1984).
\bibitem{kundrat85}V. Kundr\'{a}t and  M.V. Lokajic\u{e}k, Phys. Rev. 
D{\bf 31}, 1045 (1985); Phys. Lett B{\bf 232}, 263 (1989); Z. Phys. C
{\bf 63}, 619 (1994); Phys. Rev. D{\bf 55} (1997) (to be published).
\bibitem{kawasaki96}M. Kawasaki, T. Maehara, and M. Yonezawa, Phys. 
Rev. D{\bf 55} (1997) (to be published).
\bibitem{martini95}A.F. Martini and M.J. Menon, in {\it 1994 XV 
Brazilian National Meeting on Particles and Fields}, Angra dos Reis, 
RJ, 1994 (Sociedade Brasileira de F\'{\i}sica, S\~ao Paulo, 
1995) p. 208.
\bibitem{bronzan74}J.B. Bronzan, G.L. Kane and U.P. Sukhatme, Phys. 
Lett. B{\bf 49}, 227 (1974).
\bibitem{ayres77}Differential cross section data at 
$\protect\sqrt{s}=13.8$ and 19.4 GeV: D.S. Ayres {\it et al}., 
Phys. Rev. D{\bf 15}, 3105 (1977); C.W. Akerlof {\it et al}., 
Phys. Rev. D{\bf 14}, 2864 (1976); G. Fidecaro {\it et al}., 
Phys. Lett. B{\bf 105}, 309 (1981); R. Rubinstein {\it et al}., 
Phys. Rev. D{\bf 30}, 1413 (1984).
\bibitem{fajardo81}$\rho$ data at $\protect\sqrt{s}=13.8$ and 
19.4 GeV: L.A. Fajardo {\it et al}., Phys. Rev. D{\bf 24}, 46 (1981).
\bibitem{augier93}C. Augier {\it et al}., Phys. Lett. B{\bf 316}, 
448 (1993).
\bibitem{gotsman93}E. Gotsman, E.M. Levin, and U. Maor, Z. Phys. C
{\bf 57}, 677 (1993).
\bibitem{cheng87}H. Cheng and T.T. Wu, {\it Expanding protons:
scattering at high energies} (MIT Press, Cambridge, MA, 1987).
\bibitem{bevington92}P.R. Bevington and D.K. Robinson, {\it Data 
reduction and error analysis for the physical sciences}, (McGraw-Hill
, New York, 1992).
\bibitem{martinime96}A.F. Martini and M.J. Menon, in {\it 1995 XVI 
Brazilian National Meeting on Particles and Fields}, Caxambu, 
MG, 1995 (Sociedade Brasileira de F\'{\i}sica, S\~ao Paulo, 
1996) p. 305.
\bibitem{carrol76}A.S. Carrol {\it et al}., Phys. Lett. B{\bf 61}, 
303 (1976); A.S. Carrol {\it et al}., Phys. Lett. B{\bf 80}, 423 
(1979).
\bibitem{giromini94}P. Giromini {\it et al}. (CDF Collaboration), in 
{\it Proceedings of the Fifth International Conference on Elastic and 
Diffractive Scattering}, edited by H.M. Fried, K. Kang, and C.-I. 
Tan (World Scientific, Singapore, 1994), p. 30.
\bibitem{lukaszuk73}L.Lukaszuk and B. Nicolescu, Nuovo Cimento Lett. 
{\bf 8}, 405 (1973); P. Gauron, E. Leader and B. Nicolescu, Phys. 
Rev. Lett. B{\bf 54}, 2656 (1985); B{\bf 55}, 639 (1985).
\bibitem{giffon96}M. Giffon, E. Predazzi, and A. Samokhin, Phys. Lett. 
B{\bf 375}, 315 (1996).
\bibitem{chou67}T.T. Chou and C.N. Yang, in {\it High Energy Physics 
and Nuclear Structure}, edited by G. Alexander (North-Holland, 
Amsterdam, 1967), p. 348; Phys. Rev. {\bf 170}, 1591 (1968); Phys. 
Rev. Lett. {\bf 20}, 1213 (1968).
\bibitem{henzi83}R. Henzi and P. Valin, Phys. Lett B{\bf 132}, 443 
(1983); Phys. Lett B{\bf 160}, 167 (1985).
\bibitem{amaldi76}U. Amaldi, M. Jacob, and G. Matthiae, Ann. Rev. 
Nucl. Sci. {\bf 26}, 385 (1976).
\bibitem{glauber84}R.J. Glauber and J. Velasco, Phys. Lett. B 
{\bf 147}, 380 (1984); in {\it Proceedings of the Second 
International Conference on Elastic and Diffractive Scattering}, 
edited by K. Goulianos (Editions Fronti\`eres, Gif-Sur-Yvette, 1988), 
p. 219.
\bibitem{kamran88}M. Kamran and I.E. Qureshi, in {\it Proceedings of 
the Second International Conference on Elastic and Diffractive 
Scattering}, edited by K. Goulianos (Editions Fronti\`eres, 
Gif-Sur-Yvette, 1988), p. 289; Hadronic J. {\bf 12}, 25 (1989); 
{\bf 12}, 173 (1989).
\bibitem{saleem88}M. Saleem, F. Aleem, and I.A. Azhar, Europhys. Lett. 
{\bf 6}, 201 (1988); F. Aleem, M. Saleem and G.B. Yodh, J. Phys. G 
{\bf 16}, L269 (1990).
\end{references}
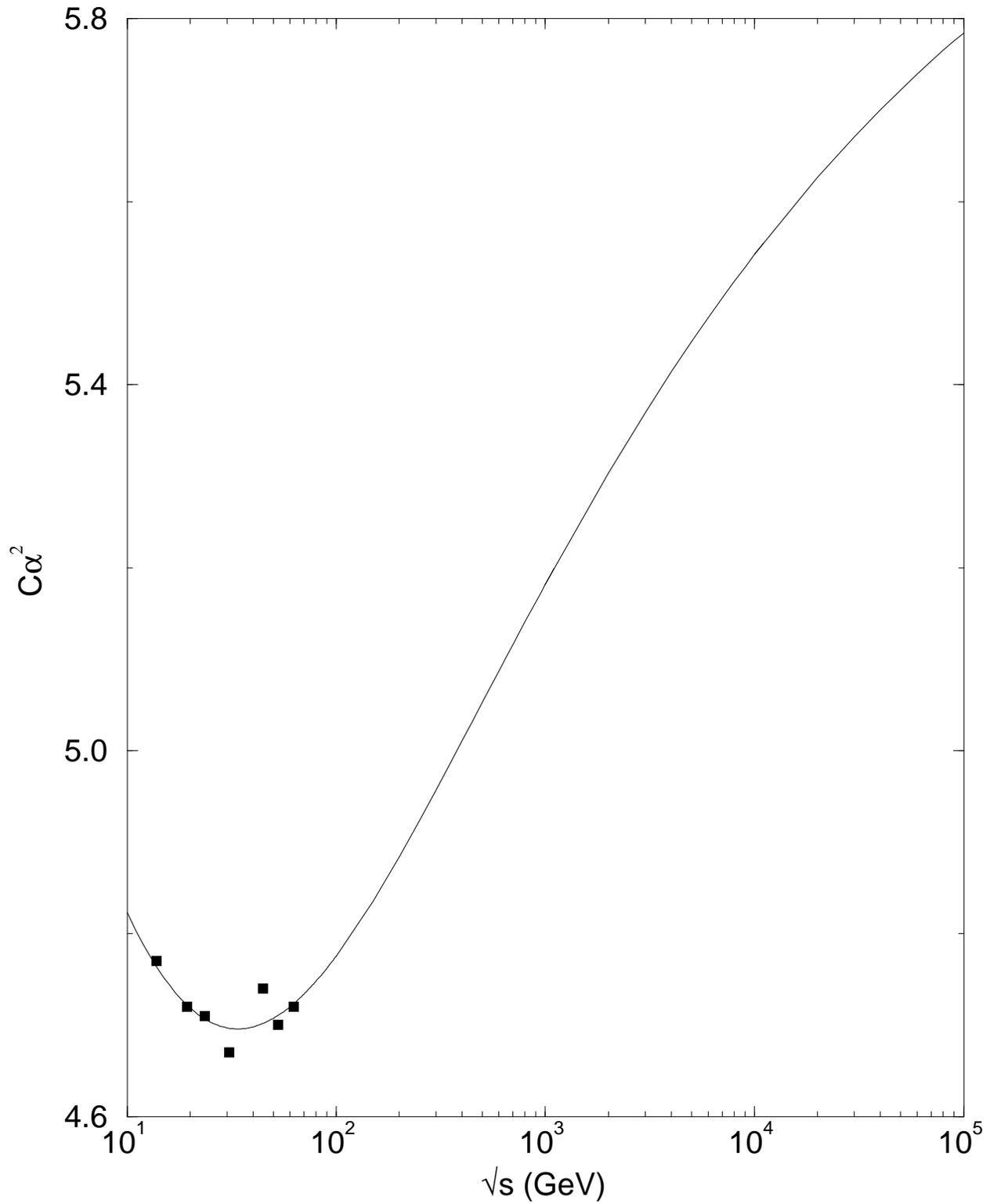
\end{document}